\documentclass[10pt]{article}

\usepackage{fullpage}

\usepackage{graphicx}
\usepackage{color,graphics,amssymb,amsthm,amsmath}
\usepackage{caption,subcaption} 
\usepackage{rotating}
\usepackage{authblk}  
\usepackage{textcomp} 

\newcommand{\argmin}[1]{\underset{#1}{\operatorname{arg}\!\operatorname{min}}\;}
\newcommand{\pg}{\color{black}}

\makeindex

\begin{document} 
\title{Let the paintings play}
\author[1]{Paola Gervasio}
\author[2]{Alfio Quarteroni}
\author[3]{Daniele Cassani}
\affil[1]{DICATAM, Universit\`a degli Studi di Brescia, Brescia (Italy)}
\affil[2]{MOX, Department of Mathematics, Politecnico di Milano, Milano (Italy) and Institute of Mathematics, Ecole Polytechnique Federale de Lausanne, Lausanne (Switzerland) (professor emeritus)}
\affil[3]{DiSAT, Universit\`a degli Studi dell'Insubria and Riemann International School of Mathematics, Varese (Italy)}

\maketitle
\begin{abstract}
In this paper we introduce a mathematical method to extract similarities between paintings and music tracks. Our approach is based on the digitalization of both paintings and music tracks by means of finite expansions in terms of orthogonal basis functions (with both Fourier and wavelet bases). The best fit between a specific painting and a sample of music tracks from a given composer is achieved via an $L^2$ projection upon a finite dimensional subspace.
Several examples are provided for the analysis of a collection of works of art by the Italian artist Marcello Morandini. Finally we have developed an original applet that implements the process above and which can be freely downloaded from the site \texttt{https://github.com/pgerva/playing-paintings.git}
\end{abstract}




\section{Introduction}
Art expresses itself through a surprisingly wide variety of languages: music and paintings represent two extraordinary examples.

Despite their striking diversity, painting and music can sometimes evoke
similar sensations. We can appreciate beautiful music by clearing our mind and imagine landscapes that inspire us, or surprising ourselves while admiring a beautiful painting to think of a musical composition that this painting evokes.

It goes without saying that this feeling of transport \emph{from one world to
another} (that of painting and that of music) is completely subjective.
Different people will experience it in a completely
different way. This too, after all, is a form of art, the way we free our mind
in search of unique emotional moments. {\pg In this paper, however, we want to
follow a more objective path.}

{\pg Can we try to interpret this transfer} with the help of mathematics? By making it
more rigorous (and also less arbitrary), confining it within a more objective
interpretative setting guided by mathematical criteria?
This is the question we will try to partially answer in this paper. 'One' answer among the many possible, since the metric that we will introduce to establish this possible analogy between two very different worlds (painting and music) will inevitably condition our answer.

More precisely, the specific question we will try to answer is the
following: 'Can we {\pg reproduce} a painting from the musical point of view?'

There are obviously many ways to do it, and different artists {\pg and
scientists} have given their seminal contribution to this matter in the past.

At first, let us mention UPIC (Unit\'e Polyagogique Informatique CEMAMu) of
Xenakis, an electronic system composed by a graphical whiteboard, a computer, a
digital/analogical converter and an acoustic system, designed to transform two--dimensional hand made drawings into sounds \cite{Xenakis, Mazzola-2016}. The translation into music by UPIC is carried out by associating graphical signs to specific musical parameters. 

{\pg Mazzola used module
theory, category theory, homotopy theory, and algebraic geometry to analyse
Beethoven's ``Hammerklavier'' Sonata op. 106, and to compose
a new sonata from this analysis. Moreover, thanks to 
the graphical composition software
\emph{presto}\textsuperscript{\textregistered}, he composed
a concert for piano and percussion
\cite{Mazzola-The-topos-2002, Mazzola-2016, Mazzola-synthesis}. 
}

Another interesting approach has been designed by Mannone \cite{Mannone-2011, Mannone-2017}. Starting from the idea of the Sicilian musician Betta of representing a musical score in the 3D space (Intensity, Onset, Pitch), Mannone took an inverse approach by 'musicalizing' a three-dimensional object by projecting it inside the (Intensity, Onset, Pitch) space, and by choosing appropriately the level of discretisation (how close are the points of the image to be considered as musical notes), the range of pitches, times and duration of sounds. 

In \cite{Mannone-2022}, the Quantum GestART \cite{Mannone-gestart} and the gestural similarity conjecture \cite{Mannone-2018} were applied to mathematically analyse and compare patterns and structures of Qwalala (an installation by Pae White realized for the Venice Biennale), with an ensemble musical piece derived from it.

{\pg A parallel line of research has moved recently in the field of multimedia,
signal theory, and machine learning.
We mention, e.g., \cite{Payling-2014}
and  \cite{Santini-2019}, who propose to 
generate sounds according to image analysis and colour detection, respectively.
In \cite{Zhou-2018}, the authors apply learning--based methods to generate raw
waveform samples, starting from given video frames, while in \cite{Zhao-2020}, machine
learning algorithms are used to study end--to--end matching between image and
music, based on emotions in the continuous valence--arousal space
\cite{Tsiourti-2019}.}

The approach that we will pursue in this paper is a 
\emph{feature extraction process} to find similarities (more precisely similarity features)
between images and music tracks. 
{\pg This is a rigorous mathematical approach inspired by the arts.}

We will apply our approach to a sample of paintings of maestro
Marcello Morandini
\cite{Morandini-catalogo2020,Morandini-fondazione2021,Morandini-wikipedia2021} and will let
every considered painting 'play' according to some preselected music tracks
from a few extraordinary classical and jazz composers. We want however make
clear that our mathematical approach is very general, it is fully automatic, and can be applied to any
other possible combination of painters and music composers.

We proceed in the following way. For each considered painting, we will
\emph{expand it} with respect to a basis of orthonormal functions.
Namely, we consider both the Fourier basis and the wavelet basis (with
different types of wavelets). Nowadays, Fourier and wavelets systems are the
most widely used tools in the field of digital image and sound processing for
digital/analog conversion, signals compression and manipulation {\pg (far from
being exhaustive, see, e.g.,
\cite{Mallat1998, GonzalezWoods2008, Su-2001, Li-2004,
Peeters-2011,Cemgil-2003,Alm-2002,Lang-1998,Kronland-Martinet-1988,Ishi-2009,Sharma-2020,Alias-2016} and the
references therein)}. Their broad
diffusion is also due to the existence of very fast transforms
\cite{CooleyTukey1965, Mallat1998, daubechies1992} that allow to manage heavy
computations in real time.

After having {\pg loaded} the {\pg waveform of the} music track of the various
reference musical works considered, we {\pg compute} for each {\pg track} the
development with respect to the same basis (Fourier's or wavelets'). At this
point, we generate the finite dimensional space of the musical works and
project (in the sense of $L^2$) the finite development of the painting upon
this subspace. We consider this result (of best approximation) as the best
possible representation of the
picture we have examined, in the context of the music tracks considered.

Next, we introduce {\pg different metrics to evaluate our results. From one
hand  we measure the distance between the new piece of music generated by the
algorithm and the space spanned by the original piece of music of a specific
composer. On the other
hand we measure the similarity between each original sound track and the
chosen painting, for what concerns intrinsic features.}

Finally, we have created a Python app that, {\pg starting from any database of
paintings and pieces of music, allows the user} to automatically search for similarities in the sense specified above.
This app can be freely obtained from the github repository \cite{pgerva-github-playing-paintings2022}.

The summary of the paper is as follows. In Section \ref{sec:audio} we recall the fundamentals of Fourier and wavelet transforms for audio signals, while Section \ref{sec:images} is devoted to images transforms. The projection approach is described in Section \ref{sec:minquad} and the python applet is described in Section \ref{sec:app}. In Section \ref{sec:numres} we apply the python app and make a similarity analysis among some operas of maestro Marcello Morandini and  several music tracks from classical and jazz repertoire.
{\pg Finally, Section \ref{sec:conclusions} recaps the conclusions and anticipates future
developments of this work.}

\section{The audio signal and its transforms}\label{sec:audio}
From the mathematical point of view, a digital audio track is a real one--dimensional function $s(t)$ that is sampled at a set of equally spaced points belonging to a bounded interval $\overline\Omega=[0,T]$, with $T>0$.
To analyse the digital audio track, the function (or signal) $s(t)$ is transformed 
into the so-called space of frequencies.

\subsection{The Fourier transform}
Let $L^2(\Omega)=\{v:\Omega\to {\mathbb
R}: \ \int_\Omega v^2 <\infty\}$ be the space of functions  whose 
square is summable in $\Omega$ according to the Lebesgue measure, equipped with the inner product
$(u,v)=\int_\Omega u\, v$ and the induced norm $\|v\|=(v,v)^{1/2}$.

Let us introduce the set of functions $\varphi_k(t)=e^{i\frac{2\pi}{T} kt}$,  with $k\in{\mathbb Z}$ and $i$ the imaginary unit, that forms an orthogonal basis of  $L^2((0,T))$. 
If the audio signal $s(t)$ is a function in $L^2({\mathbb R})$, then we can  write its Fourier expansion
\begin{equation}\label{IFT}
 s(t)=\sum_{k=-\infty}^{+\infty} s_k \varphi_k(t)= 
 \frac{a_0}{2}+\sum_{k=1}^{+\infty}\left(a_k\cos\left(\frac{2\pi}{T} kt\right)
+b_k\sin\left(\frac{2\pi}{T} kt\right)\right), 
 \end{equation}
 where 
 \begin{equation}\label{FT}
     s_k=\frac{(s,\overline{\varphi}_k)}{\|\varphi_k\|^2}=\frac{1}{T}\int_{-\infty}^{\infty} s(t)e^{-i\frac{2\pi}{T} kt}dt, \qquad k\in {\mathbb Z}
\end{equation}
are the Fourier coefficients, and
\begin{equation*}
a_k= s_k+ s_{-k},\qquad b_k=i(s_k- s_{-k}), \qquad \forall k\in {\mathbb N}.
\end{equation*}

The variable $k$ is called \emph{frequency} and the value $|s_k|$ is the amplitude of the $k-$th frequency. The map associating with $s(t)$ the sequence of its Fourier coefficients $\hat s_k$ is called \emph{Fourier transform} of $s$.

Since in practice  $s(t)$ is not often known at any point $t$ belonging to a given interval, but only at $N$ equally spaced points $t_j$ (for simplicity we fix $N$ even), the Fourier transform (\ref{FT}) is replaced by the \emph{Discrete Fourier Transform} (DWT)
\begin{equation}\label{DFT}
     \hat s_k{\pg =\frac{1}{N}}\sum_{j=0}^{N-1} s(t_j)e^{-ik {\pg j}2\pi/N},
\qquad k=-\frac{N}{2},\ldots,\frac{N}{2}{\pg -1},
\end{equation}
with the constraint that $\hat s_{-N/2}=\hat s_{N/2}$.  Starting from the discrete frequencies $\hat s_k$, we can reconstruct the signal on the whole interval $(0,T)$ by the formula
\begin{eqnarray}\label{inverse-fourier}
  {\pg s_N(t)=\sum_{k=-N/2}^{N/2-1} \hat s_k e^{ikt 2\pi/T}.}
\end{eqnarray}
In fact, $s_N(t)$ is the trigonometric interpolant of $s$ at the $N$ equally
spaced points $t_j$, i.e.,  the function $s_N\in V_N=span_{{-N/2\le k\le
N/2{\pg -1}}}\{\varphi_k\}$ satisfying the interpolation relations $s_N(t_j)=s(t_j)$ for $j=0,\ldots,N-1$. Formula (\ref{inverse-fourier}) is also known as \emph{discrete inverse Fourier transform} since, given the coefficients $\hat s_k$, it returns the values $s_N(t_j)$.

Fourier transform and its inverse are useful tools to extract features (like, e.g., the amplitudes of fundamentals harmonics) from a signal and to reconstruct it, respectively.

\begin{figure}
\begin{center}
\includegraphics[width=0.48\textwidth]{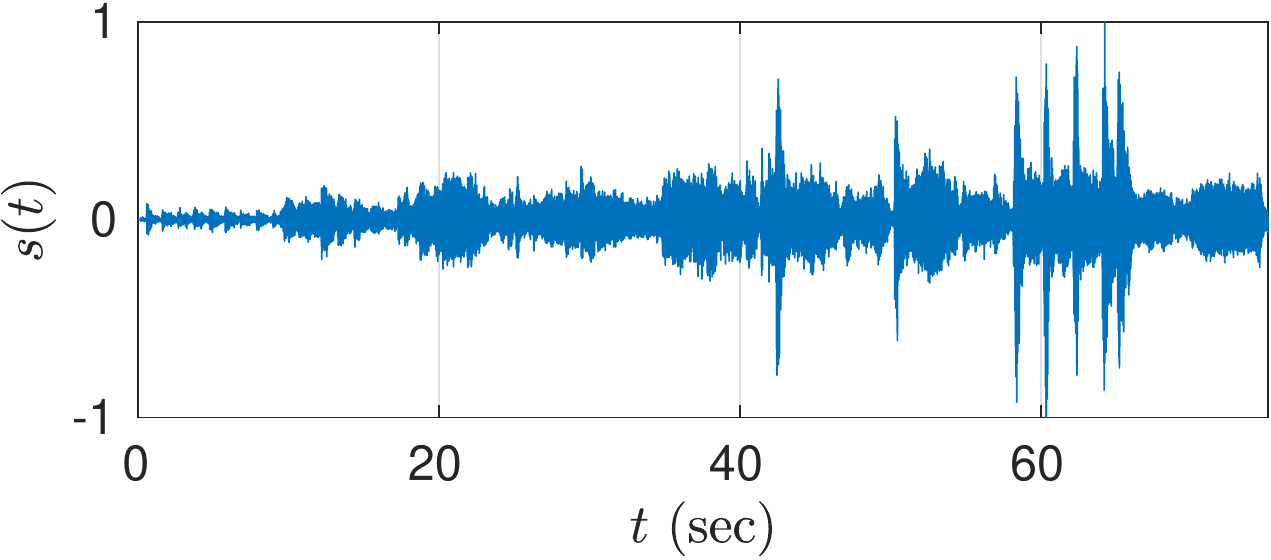}\quad
\includegraphics[width=0.48\textwidth]{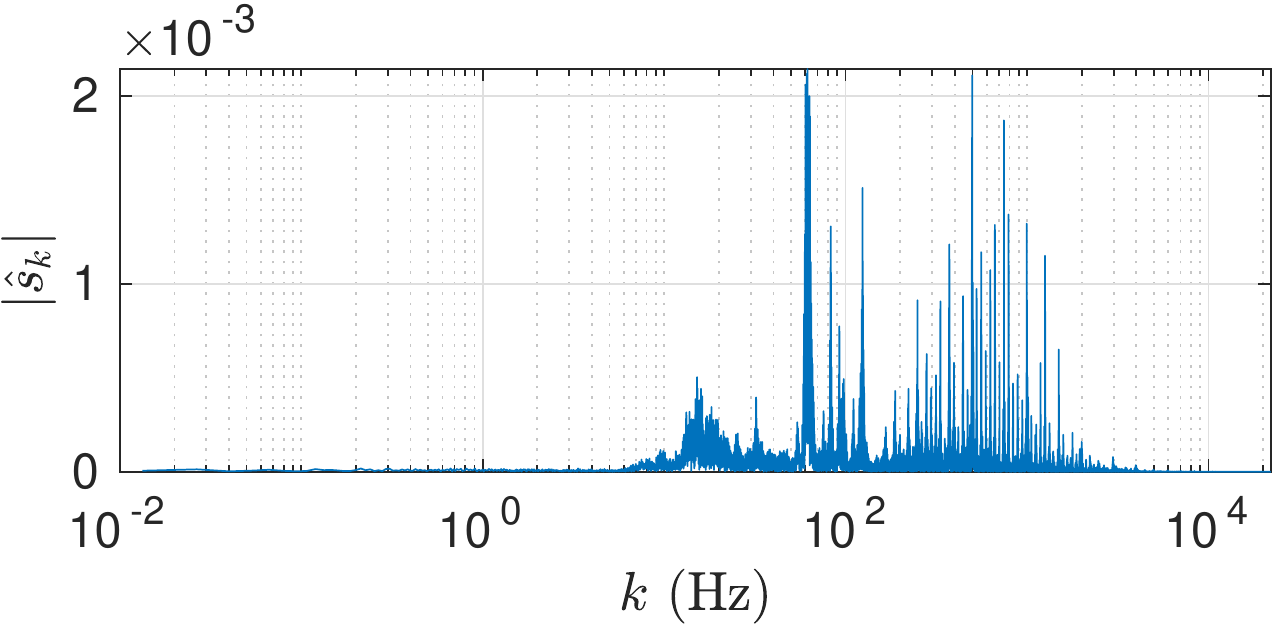}
\end{center}
\caption{The signal $s(t)$ of the 'Dance of the sugar plum fairy' in the
Nutcracker of Tchaikovsky  (on the left) and its Fourier spectrum (on the
right).} 
\label{fig:segnale-spettro}
\end{figure}

The set 
\begin{equation}\label{fourier-spectrum}
\hat{\bf s}= \{|\hat s_k| \} \in {\mathbb R}^N,
\end{equation}
with $\hat s_k$ defined in (\ref{DFT}) is the so-called \emph{(discrete) Fourier spectrum} of the signal $s(t)$ and it is usually computed by the \emph{Fast Fourier Transform} (FFT) \cite{CooleyTukey1965}. 
In Figure \ref{fig:segnale-spettro} a signal and its discrete Fourier spectrum are shown.

The Fourier transform is a widely used tool to analyse sounds and extract
features. However, it tells us which frequencies occur in the music track, but
not at which time they occur. It is like saying that we know which notes make
up a piece of music and how frequently they occur, but we do not know when
these notes {\pg are to} be played and for how long.
{\pg A remedy could be resorting to windowed 
or Short Time Fourier Transform (STFT) (see, e.g., \cite{Polikar-2003}).
However, the results of a STFT strongly depends on the size of the 
windows considered, with the drawback to loose accuracy in either time or
frequency. For this reason we have chosen to apply}
wavelet transform instead of STFT; we introduce wavelets in the next section.

\subsection{The Wavelet Transform}
 
Wavelets provide the time-frequency analysis of a signal. This means that, not only the frequencies and their amplitudes can be extracted by the signal, but also at which time a certain frequency occurs. 

Moreover, wavelets enable \emph{multi-resolution analysis}, i.e., the
possibility to analyse a signal (either a music track or an image) at many
different resolution (accuracy) levels. Thus, the features that might not
be recognized at one resolution of the signal may be easily detected to a further level of accuracy.

\begin{figure}
\begin{center}
\includegraphics[width=0.4\textwidth]{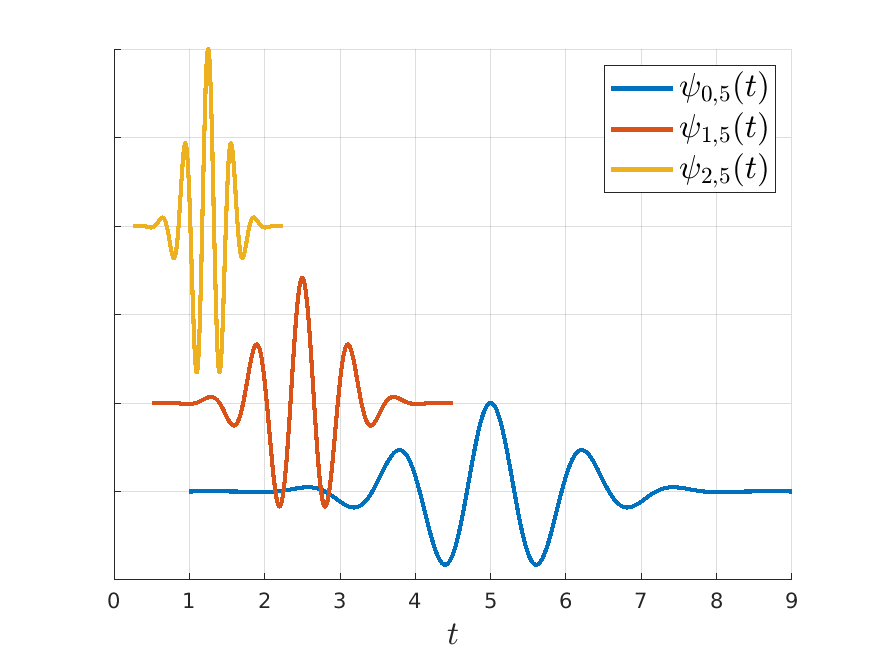}
\end{center}
\caption{Wavelets functions $\psi_{j,k}(t)$ (\ref{discrete-wavelets}) based on
the Morlet mother wavelet, at  three different scales and centered at three
different times.} 
\label{fig:wavelet1}
\end{figure}

 
We start by considering two real functions $\phi(t)$ and $\psi(t)$, with compact support, named \emph{scaling function} and \emph{mother wavelet}, respectively. Each one of these functions can be scaled and translated on the real axis by introducing two integer parameters $j$ and $k$, named \emph{scale index} and \emph{time index}, respectively and by defining
\begin{equation}\label{discrete-wavelets}
\phi_{j,k}(t)=2^{j/2}\phi(2^j t-k), \qquad 
\psi_{j,k}(t)=2^{j/2}\psi(2^j t-k).
\end{equation}
Then, for any $j$, the scaling functions space and the wavelets space are defined, respectively, by
\begin{equation}\label{wavelt-spaces}
V_j=\overline{span_k\{\phi_{j,k}\}_k},\qquad
W_j=\overline{span_k\{\psi_{j,k}\}_k}.
\end{equation}
For a fixed time index $k$, as $j$ increases, the functions $\phi_{j,k}(t)$ and $\psi_{j,k}(t)$ cover different frequency ranges. More precisely, if we consider all the scale indexes $j\ge j_0$, for a given integer $j_0\ge 0$, the scale corresponding to $j_0$ is the coarsest possible one and it is able to detect the lower frequencies of the signal; then, by increasing $j$, the frequency increases as well, up to reach the highest ones (see Fig. \ref{fig:wavelet1}).

Viceversa, for a fixed scale index $j$, as $k$ changes, the supports of the functions move on the real axis, ranging all the time interval.

The scaling function and the mother wavelet have to be chosen carefully since they must satisfy some fundamental properties in order to guarantee an efficient and accurate multi-resolution analysis of signals.

First of all, $\phi$ and $\psi$ need to guarantee that, for any integer $j\geq j_0$, the sets $\{\phi_{j,k}\}_k$ and $\{\psi_{j,k}\}_k$ are two orthonormal systems in $L^2({\mathbb R})$ .

On the one hand, the scaling function $\phi$ must enjoy the following {\pg
properties}:
 $V_j\subset V_{j+1}$ for any $j\ge j_0$ (this ensures that any scaling function $\phi_{j,k}$ can be written as a linear combination of the scaling functions at the next (higher) scale $j+1$); if $j_0=-\infty$, then the unique intersection among all the $V_j$ is the null function; and finally $\displaystyle \lim_{j\to +\infty}V_j=L^2({\mathbb R})$.

On the other hand, the mother wavelet $\psi$ needs to ensure that the wavelet
space $W_j$ spans the difference between two successive scaling function spaces $V_j$ and $V_{j+1}$, i.e.,
\begin{equation}\label{sommadiretta}
V_{j+1}=V_j\oplus W_j \qquad \mbox{for any } j\ge j_0.
\end{equation}

Thus, the space $L^2({\mathbb R})$ can be decomposed as the direct sum of one
scaling function spaces and infinite wavelet spaces as:
\begin{equation}\label{sommadirettaL2}
L^2({\mathbb R})=V_{j_0}\oplus W_{j_0} \oplus W_{j_0+1} \oplus W_{j_0+2} \oplus \ldots.
\end{equation}
It follows that any function $s\in L^2({\mathbb R})$ admits the so-called \emph{wavelet expansion}
\begin{equation}\label{wavelet-expansion}
s(t)=s_a(t)+s_d(t)=
\sum_k c_{j_0,k}\phi_{j_0,k}(t) + \sum_{j=j_0}^{+\infty}\sum_k d_{j,k}\psi_{j,k}(t),
\end{equation}
where, for any $j\ge j_0$ and $k\in {\mathbb Z}$,
\begin{equation}\label{wavelet-coefficients}
c_{j_0,k}=\int_{\mathbb R} s(t)\phi_{j_0,k}(t)dt \quad \mbox{ and }\quad 
d_{j,k}=\int_{\mathbb R} s(t)\psi_{j,k}(t)dt
\end{equation}
are the \emph{approximation} (or \emph{scaling}) \emph{coefficients} and
the \emph{detail} (or \emph{wavelet}) \emph{coefficients}, respectively, of the expansion.

The function $s_a(t)$ is the approximation of $s(t)$ in the space $V_{j_0}$, while $s_d(t)$ is the remainder $s(t)-s_a(t)$ and takes into account the high--frequency details that are not detected in the space $V_{j_0}$.

When the function $s$ is known at $N$ equally spaced points $t_n$ (with $n=0,\ldots, N-1$), we can compute the \emph{Discrete Wavelet Transform} (DWT) of the signal $s$, that is the set of the values 
\begin{equation}\label{DWT}
\hat c_{j_0,k}=\frac{1}{N} \sum_{n=0}^{N-1} s(t_n)\phi_{j_0,k}(t_n), \quad 
\hat d_{j,k}=\frac{1}{N} \sum_{n=0}^{N-1} s(t_n)\psi_{j,k}(t_n),\quad \mbox{ for } j\ge j_0,
\end{equation}
and the inverse discrete wavelet transform reads
\begin{equation}\label{inverse-wavelet}
s_N(t)=\sum_{k=0}^{2^{j_0}-1}\hat c_{j_0,k}\phi_{j_0,k}(t)+
\sum_{j=j_0}^{J-1}\sum_{k=0}^{2^j-1} \hat d_{j,k}\psi_{j,k}(t).
\end{equation}

The global set of the approximation and detail coefficients (\ref{DWT})
\begin{equation}\label{wavelet-spectrum}
    \hat{\bf s}=\{\hat c_{j_0,k}, \ \hat d_{j,k}\}\in{\mathbb R}^N
\end{equation}
 is also named  \emph{wavelet spectrum}.

Provided we rename and sort the scaling functions $\phi_{j_0,k}$ and the wavelets $\psi_{j,k}$ appropriately, the inverse wavelet transform (\ref{inverse-wavelet}) can also be written in the more compact form
\begin{equation}\label{inverse-wavelet-bis}
s_N(t)=\sum_{k=0}^{N-1}\hat s_k \varphi_k(t)
\end{equation}
that resumes the Fourier transform (\ref{inverse-fourier}) and that will be useful in our next analysis.

Typically, the total number $N$ of samplings is selected to be a power of 2 and the total number $J$ of scales that we can take in (\ref{inverse-wavelet}) is $J=\log_2 N$. When $N$ is not a power of 2, then the sampling set is padded with zeros up to the next power of 2. 
These parameters depend on the available resolution of the audio track, and the choice of $j_0$ depends on how many scales (starting from the finest one and going down to the coarser ones) one wants to represent.

As for the Fourier transform, also a \emph{Fast Wavelet Transform} is available for computing the approximation and detail coefficients (\ref{DWT}), see, e.g., \cite{Mallat1998,GonzalezWoods2008}.

Many different  mother wavelets $\psi$ (and associated scaling functions $\phi$) satisfying the aforementioned properties are known in literature \cite{daubechies1992,Mallat1998}. Some of the most common ones are: the Haar wavelet, the Morlet wavelet, the Daubechies wavelets of different order (DB$n$ with $n=1,2,\ldots$), the Fejér-Korovkin of different order (FK$n$ with $n=4,6,8,\ldots$). Some of them are more suitable to deal with audio signals, others with images.

The choice of the mother wavelet,  as well as the choice of the total number of scales, provides a different transform (\ref{DWT}) and, then, a different analysis of the signal. 

\begin{figure}
\begin{center}
\includegraphics[width=0.48\textwidth]{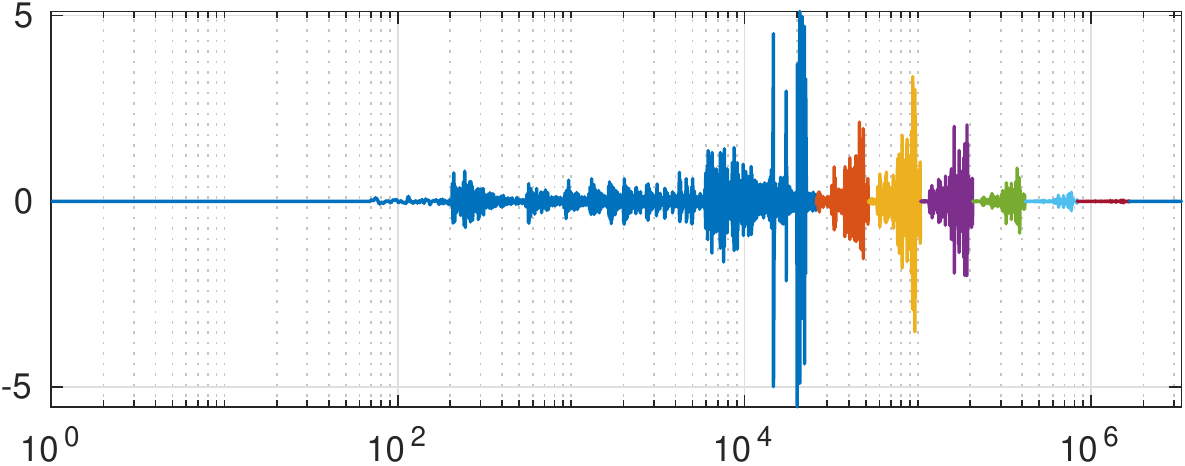}\quad
\includegraphics[width=0.48\textwidth]{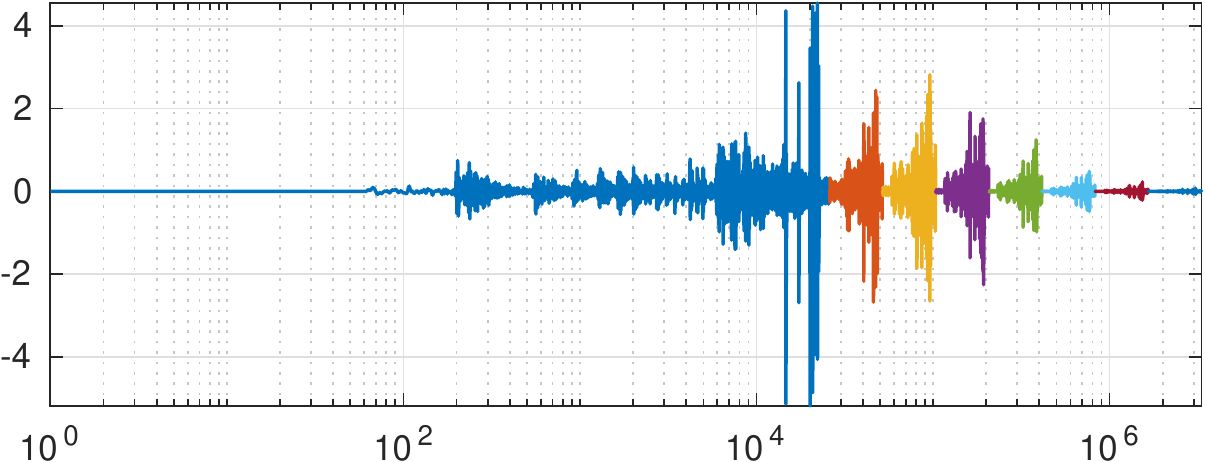}
\end{center}
\caption{Two wavelet spectra (\ref{DWT}) relative to the signal $s(t)$ plotted
in Fig. \ref{fig:segnale-spettro}, computed with Daubechies wavelets (on the
left) and  with Haar wavelets (on the right). In both cases the blue part
refers to the approximations coefficients $\hat c_{j_0,k}$, while the other
coloured parts refer to the detail coefficients $\hat d_{j,k}$, different {\pg
colours} refer to different scale indexes $j=j_0,\ldots, J$.} 
\label{fig:dwt-esempio}
\end{figure}

In Fig. \ref{fig:dwt-esempio}, two different wavelet spectra, relative to the signal $s(t)$ plotted in the left picture of Fig. \ref{fig:segnale-spettro}, are shown. 

\section{The mathematical representation of an image and its transform}
\label{sec:images}

The data of a digital image are typically stored into three matrices, each
associated with one of the three {\pg colours}: Red, Green, and Blue, so that
$[R_{ij},\ G_{ij}, \ B_{ij}]$ contain the intensity of the three {\pg colours} at the pixel $(i,j)$ of the image.
With each pixel we can associate also a unique value
\begin{equation}\label{eq:average-intensity}
p_{ij}=(R_{ij}+G_{ij}+B_{ij})/3
\end{equation}
that represents the average intensity of that pixel $(i,j)$.

{\pg When the image is stored in gray scale, the three matrices $R$, $G$, and
$B$ are replaced by a unique matrix (say, $G$), and the intensity  is defined
by $p_{ij}=G_{ij}$.}

\begin{figure}
\begin{center}
\includegraphics[width=0.26\textwidth]{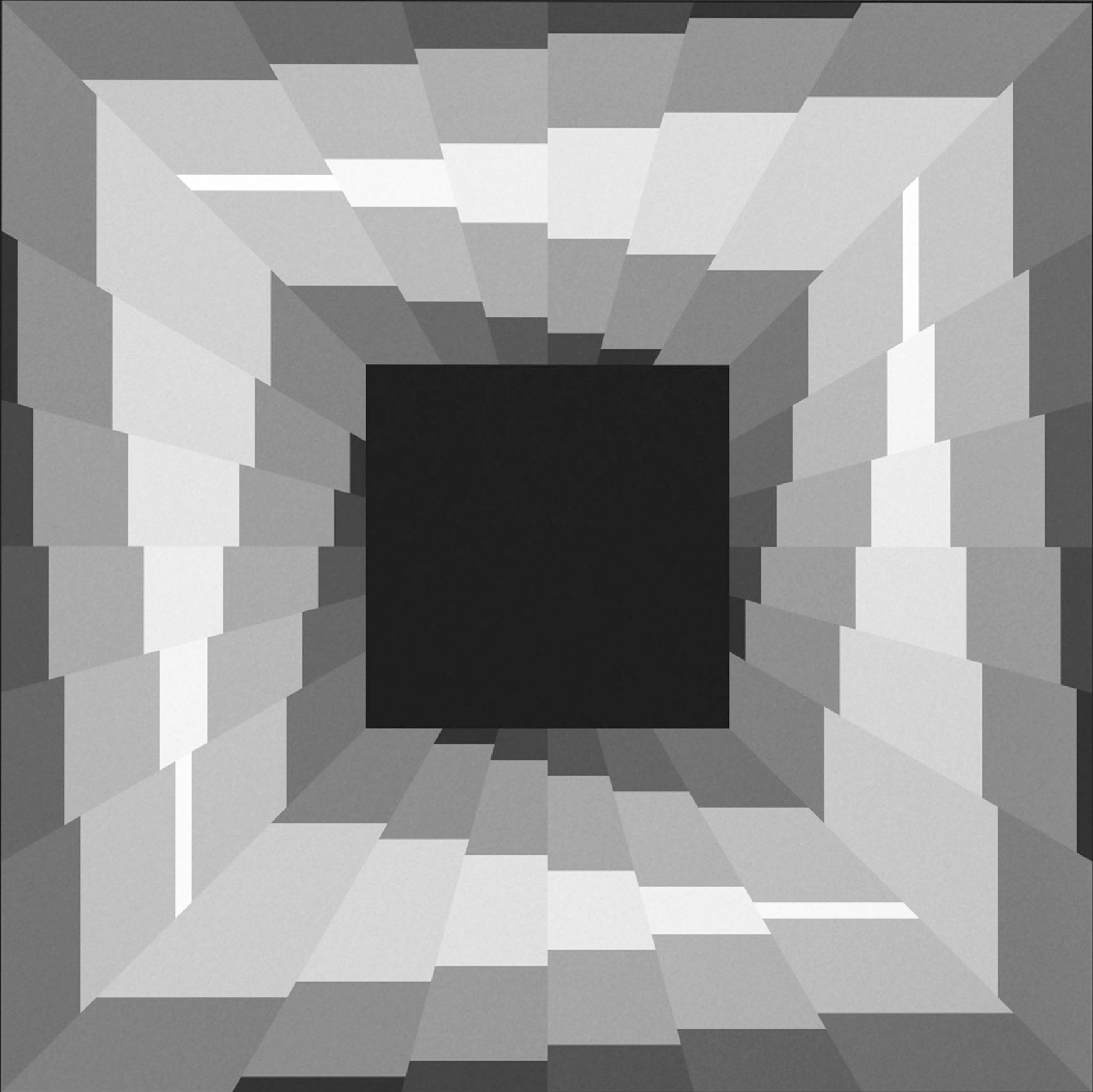}\hskip 1.cm
\includegraphics[trim=0 30 0 0, width=0.45\textwidth]{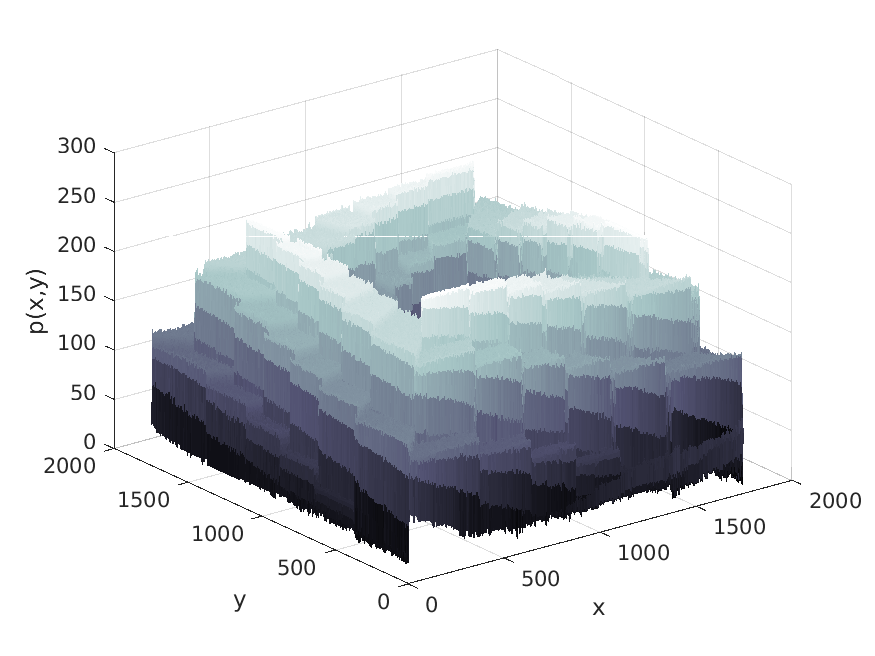}
\end{center}
\caption{Marcello Morandini's pannello 340C/1989. The digital image {\pg in
gray scale} (on the
left) and the associated intensity function (on the right).}
\label{fig:image-pixel}
\end{figure}

Since the indices $i$ and $j$ of the pixels span a bounded region of the plane,
we can interpret the average intensity as a two--dimensional function $p(x,y)$ (the variable $x$ corresponds to the column index $j$, while $y$ to the row index $i$, see Fig. \ref{fig:image-pixel}) to which we can apply suitable extensions of the discrete Fourier transform (\ref{DFT}) and of the discrete wavelet transform (\ref{DWT}).

\begin{figure}
\begin{center}
\includegraphics[width=0.3\textwidth]{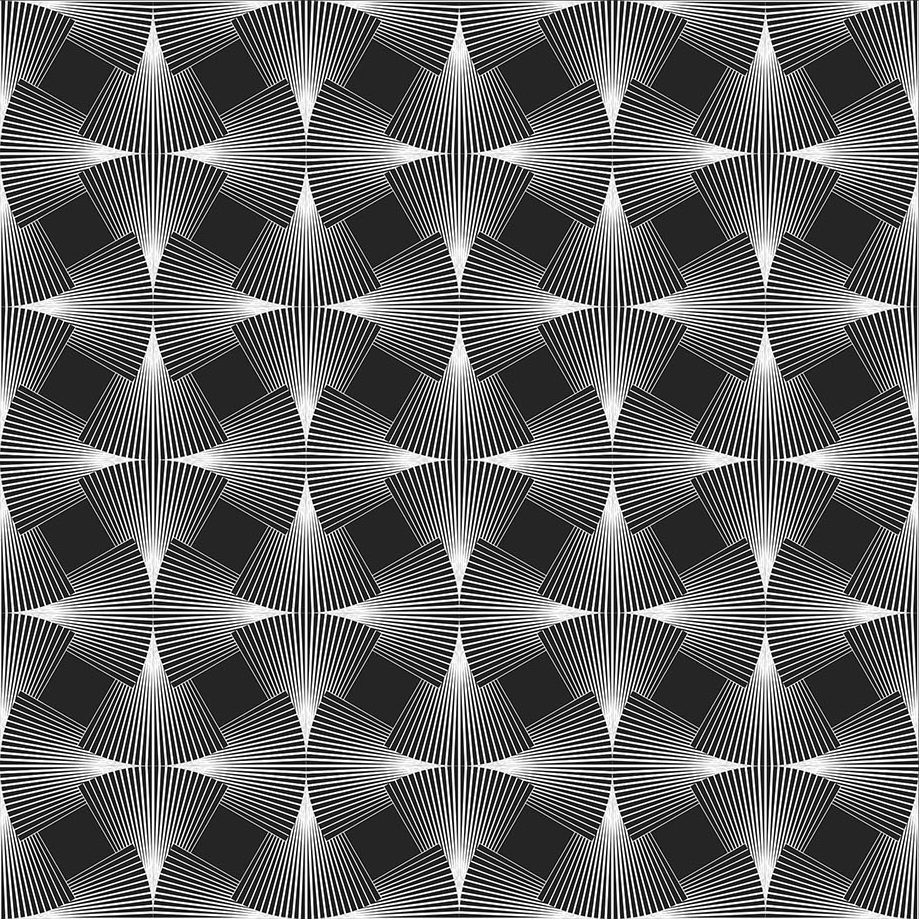}\quad
\includegraphics[width=0.3\textwidth]{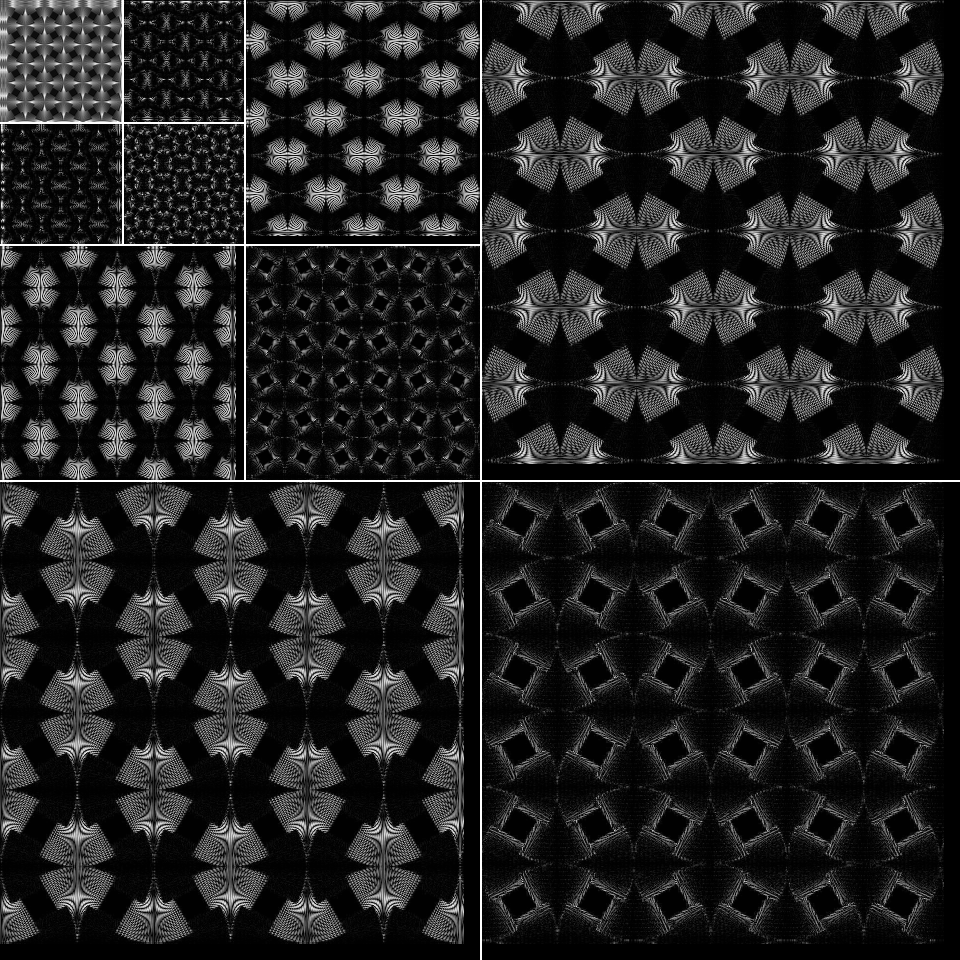}\quad
\scalebox{0.53}{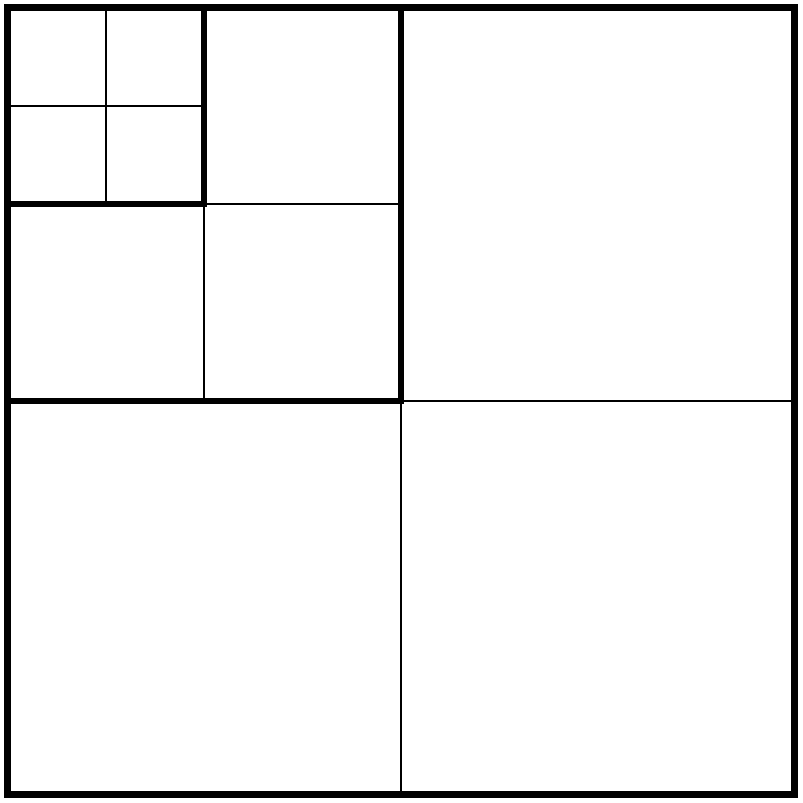}
\end{center}
\caption{Marcello Morandini's pannello 593/2012. The original image (left) and
its two--dimensional DWT transform (middle) with three scales: $j_0$, $j_0+1$ and $J=j_0+2$. On the right, the description of the coefficients shown in the middle image.}
\label{fig:2d-dwt}
\end{figure}

However, for a two--dimensional signal, there are multiple ways to  build the spectrum. In this work we have considered two approaches:
\begin{enumerate}
\item In the first one, that we name \emph{1D unrolling of the image}, we read
the pixels of the image from left to right and from bottom to top, and convert
the signal from a two--dimensional function $p(x,y)$ to a one--dimensional function $p(t)$, by interpreting the only variable $t$ as a time (with an abuse of notation we use again the same symbol $p$).

Then, we proceed by applying one--dimensional transforms as we did with the audio signals and we compute the Fourier or the wavelet spectrum $\hat{\bf p}$ of the painting, by applying  (\ref{DFT})--(\ref{fourier-spectrum}) or (\ref{DWT})--(\ref{wavelet-spectrum}), respectively, with $s$ replaced by $p$.

\item
The second approach consists {\pg of} applying the \emph{full 2D transform} to the digital image. 
The two--dimensional DFT is built as the tensor product of two one--dimensional DFT (see \cite[Ch. 4]{GonzalezWoods2008}).

The two--dimensional DWT is defined starting from the one--dimensional scaling
function $\phi$ and  mother wavelets $\psi$ and by defining one two--dimensional
scaling function $\Phi(x,y)=\phi(x)\phi(y)$ and three two--dimensional mother
wavelets $\Psi^H(x,y)=\psi(x)\phi(y)$, $\Psi^V(x,y)=\phi(x)\psi(y)$ and
$\Psi^D(x,y)=\psi(x)\psi(y)$ that correspond to the variations along the
horizontal, vertical and diagonal direction, respectively. Then, the
approximation coefficients $c_{j_0,k}$ and the detail coefficients $d_{j,k}^H$,
$d_{j,k}^V$,  and $d_{j,k}^D$, for any scale $j=j_0,\ldots, J$ are defined
similarly to the one--dimensional case. In Figure \ref{fig:2d-dwt}, we show a
digital image and its two--dimensional DWT computed over three scales. 
We refer, e.g., to \cite[Sect. 7.5]{GonzalezWoods2008} for a more indepth description of 2D DWT.

Finally, we end up with a Fourier or a wavelet spectrum which we call $\hat{\bf p}$.
\end{enumerate} 


Our next purpose will consist in comparing the spectrum of a digital image with the spectrum of one or several music tracks. To this aim we have to ensure that the sizes of the spectrum of the image and those of the music tracks do coincide.

One way is to transform an image of $N=N_x\times N_y$ pixels by following the first approach (1D unrolling) and then sampling a music track exactly at $N$ points.
Another suitable approach is to apply the Fourier transform. In this case the size of the discrete Fourier spectrum equals the total number of pixels of the original image and then that of the digital audios.

Instead, when we apply the full two--dimensional wavelet transform, an inconsistency between the size of the discrete spectrum and that of the original image (and consequently the size of the audio spectra) can occur. To overcome this drawback, we align the number of approximation coefficients of the image with that of the music tracks, by padding with zero values or by replicating the coefficients of the shorter sequence. Similarly we align the number of the detail coefficients at each scale of the DWT. 
This is another aspect of arbitrariness that influences the final result.

\section{Projection of the painting onto the music subspace}
\label{sec:minquad}

Let us consider an image of $N$ pixels ($N=N_x\times N_y$) whose intensity function is denoted by $p(x,y)$, and 
$M\geq 2$ digital audios (corresponding to as many music tracks), whose signals are $s^{(i)}(t)$ (with $i=1,\ldots, M$) sampled at $N$ equally spaced points. For example, in our tests we have fixed $M=4$.

Then, let us choose to work with one of the two transforms: DFT or DWT, the
same for both the image and the audios. Then, let us proceed as follows:

\begin{enumerate}
\item Compute the spectrum $\hat{\bf s}^{(i)}=[\hat s^{(i)}_{k}]\in {\mathbb R}^N$  of the $i-$th audio signal $s^{(i)}(t)$; 
%
\item Let 
$$V_N=span\{\hat{\bf s}^{(1)},\ \hat{\bf s}^{(2)},\ldots,\ \hat{\bf s}^{(M)}\}\subset {\mathbb R}^N$$ 
be the \emph{music subspace}, i.e., the space spanned by the spectra of the audio signals; 
\item Compute the spectrum $\hat{\bf p}\in {\mathbb R}^N$ of the image; 
%
\item Look for
\begin{equation}\label{least-squares}
\widetilde{\bf s}=\argmin{\hat{\bf s}\in V_N}\|\hat{\bf p}-\hat{\bf s}\|_2^2,
\end{equation}
that is look for the least squares projection of the spectrum of the painting
onto the music subspace $V_N$ with respect to the euclidean norm $\|\cdot\|_2$. The solution $\widetilde{\bf s}$ is named \emph{best approximation spectrum} of the picture in the music subspace.
Since $\widetilde{\bf s}\in V_N$, it can be written as
\begin{equation}\label{expansion-sol-ls}
\widetilde{\bf s}=\displaystyle \sum_{i=1}^M
\widetilde\alpha_i\hat{\bf s}^{(i)}
\end{equation}
with
$\widetilde \alpha_i\in {\mathbb R}$. It follows that the least squares problem (\ref{least-squares}) can be written equivalently as: look for the array $\widetilde{\boldsymbol\alpha}$ solution of 
\begin{equation}\label{least-squares-alpha}
\widetilde{\boldsymbol\alpha}=\argmin{{\boldsymbol \alpha}\in {\mathbb R^M}}
\|\hat{\bf p} -\sum_{i=1}^M\widetilde\alpha_i \hat{\bf s}^{(i)}\|_2^2;
\end{equation}
\item Generate the new audio signal
\begin{equation}
\widetilde s(t)=\sum_k \widetilde s_k \varphi_k(t),
\end{equation}
where $\widetilde s$ is the \emph{best approximation music track} of the given painting.
\end{enumerate}

Clearly, $\widetilde s$ depends on which (and how many) music tracks we have chosen to build the subspace $V_N$, on the chosen transform (Fourier or wavelet) and, last but not least, on the approach used to compute the transform of the image (1D unrolling or full 2D).

\medskip
The $i-$th component $\widetilde\alpha_i$ indicates how much the spectrum of the painting 'resembles' the spectrum of the $i-$th music track.
The greater the ratio $\overline{\alpha}_i=|\widetilde\alpha_i|/(\sum_i|\widetilde \alpha_i|)$,  the greater the similarity between the spectrum of the painting and the spectrum of the $i-$th music track.
Consequently, if we wanted to choose one of the $M$ music tracks as the best representative of the painting, this would clearly be associated with the coefficient $\widetilde \alpha_i$ having the maximum modulus.

From the most qualitative point of view, our procedure identifies the best possible combination of the chosen $M$ music tracks, that provides the best similarity between the spectrum of the painting and the music subspace.

In order to quantify this similarity, we can compare the best
approximation spectrum $\widetilde{\bf s}$  with the original spectrum  $\hat{\bf s}$ of the image, by defining the distances (either absolute and relative)
{\pg 
\begin{equation}\label{distanze}
d=\left\|\frac{\widetilde{\bf s}}{\|\widetilde{\bf s}\|}-\frac{\hat{\bf
p}}{\|\hat{\bf p}\|}\right\|, \qquad \mbox{ and } \qquad
d_r=\frac{d}{\|{\bf p}\|},
\end{equation}
}
where ${\bf p}$ is the array containing the {\pg values $p_{ij}$  of the
intensity} at the pixels of the image (see (\ref{eq:average-intensity})).
The smaller the distance $d$ (or the relative one $d_r$), the greater the
similarity between the picture and the linear combination
{\pg (\ref{expansion-sol-ls})} of the $M$ chosen
music tracks.

\section{The Python applet}
\label{sec:app}
In order to make the whole process, described above, easy to use  by
people who are completely unaware of the underlying mathematics we have
designed and produced a Python app based on PySide2 (the official Python module
from  \cite{pyside2}, \cite{fitzpatrick}) that can be downloaded from the github repository \texttt{https://github.com/pgerva/playing-paintings.git}.

\begin{sidewaysfigure}
\includegraphics[width=\textwidth]{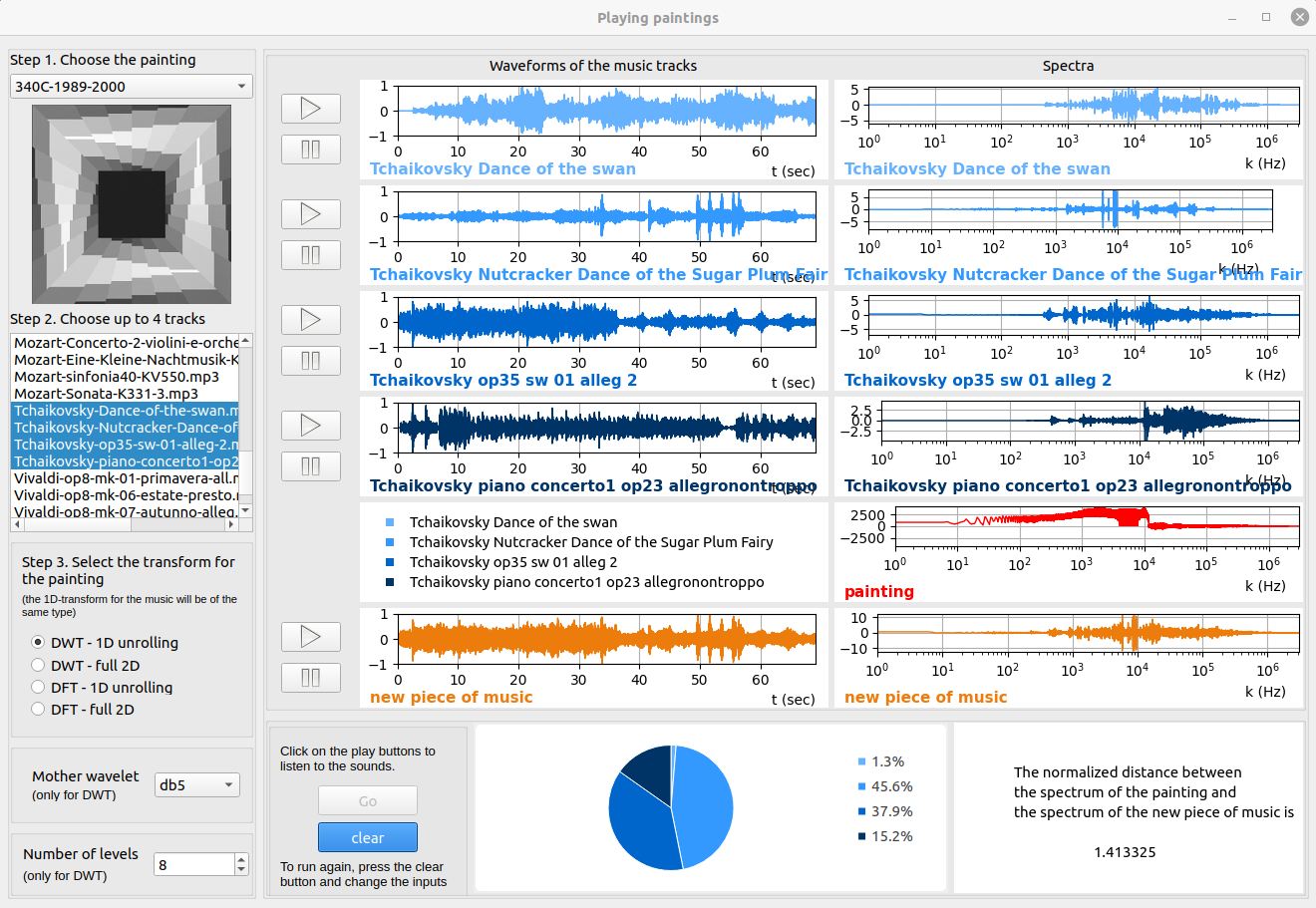}
\caption{The dashboard of the app.}
\label{fig:app}
\end{sidewaysfigure}

As an instance, in Figure \ref{fig:app} we report the app dashboard, with the processing
results on Marcello Morandini's 'pannello 340C/1989'. For instance, in order to generate the
music subspace $V_N$ we have chosen four musics by {\pg Pyotr Ilyich Tchaikovsky}, and finally
we have selected the wavelet transform with the 1D unrolling approach to elaborate the image. 

The  coefficients $\widetilde \alpha_i$ solution of
(\ref{least-squares-alpha}) are reported (as percentages) in the pie chart at
the bottom of the panel. The blue graphs in the middle part of the panel
represent the {\pg waveform of the} signals $s^{(i)}$ {\pg associated with}
the four music tracks, while the orange graph is the {\pg waveform of the} best approximation music track $\widetilde{s}(t)$ of the painting.

The graphs on the right show the wavelet spectra in the following order: first
the spectra of the four music tracks are shown in blue, then the spectrum of
the painting (in red) follows, 
while the best approximation spectrum (in orange) is on the bottom. In this
simulation we have applied the DWT with 1D unrolling of the image based on
Daubechies wavelets db5, with {\pg 8} scales. 

{\pg We notice that the waveform of the new music  differs from the
waveforms of the original sound tracks, although it shows up some similarities with the
waveforms of the second and third music tracks. Such similarities can also be 
observed  by inspecting
the weights $\widetilde{\alpha}_i$ of the formula
(\ref{expansion-sol-ls}) and, consequently, in the percentages $\overline{\alpha}_i$
shown in the legend of the pie chart. The
algorithm has computed  the weights
$\widetilde{\alpha}_1=-0.044$,
$\widetilde{\alpha}_2=-1.560$, 
$\widetilde{\alpha}_3=1.298$, and 
$\widetilde{\alpha}_4=-0.520$, 
and the percentages 
$\overline{\alpha}_1=1.3\%$,
$\overline{\alpha}_2=45.6\%$, 
$\overline{\alpha}_3=37.9\%$, and 
$\overline{\alpha}_4=15.2\%$,
showing that the weights of both the second and
the third music tracks are  bigger than those corresponding to the first and fourth audios.}

\clearpage
\newpage
\section{Similarity analysis}
\label{sec:numres}

For each painting of Marcello Morandini, we have computed the best
approximation music track with respect to the four pieces of the nine musical
composers. Then we have computed the distances $d_r$ defined in
(\ref{distanze}) and we have ordered  {\pg such distances in ascending} order and
{\pg associated} the {\pg colours} from blue (corresponding to the smallest distance
{\pg and then to the best similarity}) to
yellow (largest distance and then worst similarity).

To transform the image, we have considered both Fourier (DFT) and wavelet (DWT)
transforms with either 1D unrolling and full 2D {\pg strategies}.

\begin{figure}
\begin{center}
\includegraphics[width=0.45\textwidth]{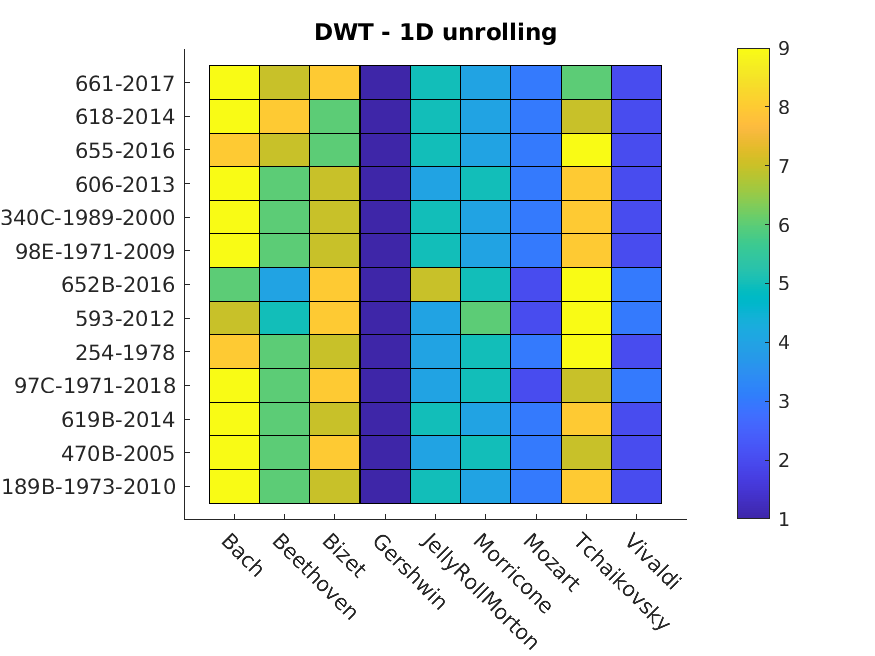}
\includegraphics[width=0.45\textwidth]{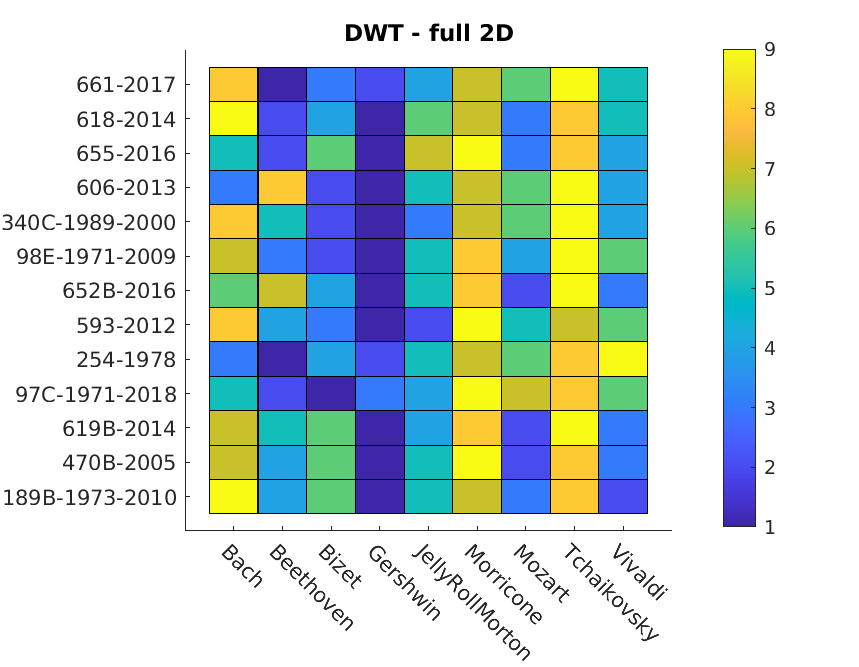}\\
\includegraphics[width=0.45\textwidth]{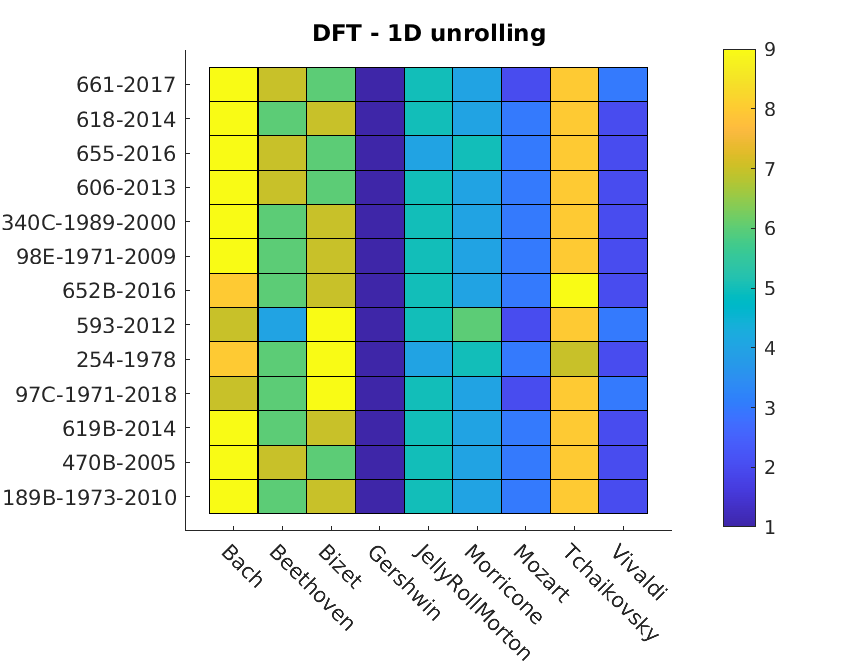}
\includegraphics[width=0.45\textwidth]{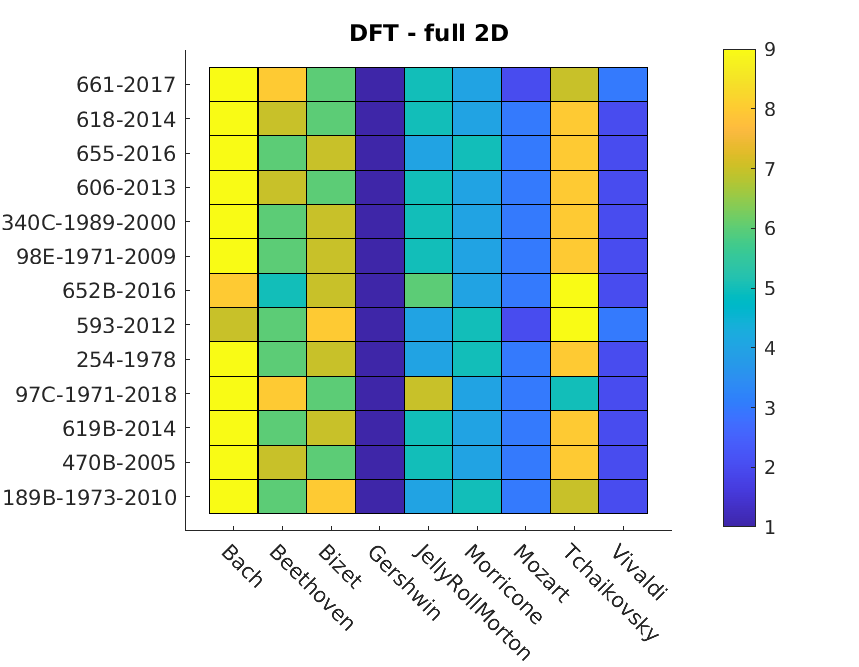}
\end{center}
\caption{Similarity matrices between Morandini's paintings and composers' music, the dark blue {\pg
colour} corresponds
to a greater similarity, the yellow to a lesser similarity. {\pg The top--right
picture (DWT with full 2D transform) has been obtained by adopting the strategy
of replicating the coefficients of the image's transform}}
\label{fig:sintesi1}
\end{figure}

\begin{figure}
\begin{center}
\includegraphics[width=0.45\textwidth]{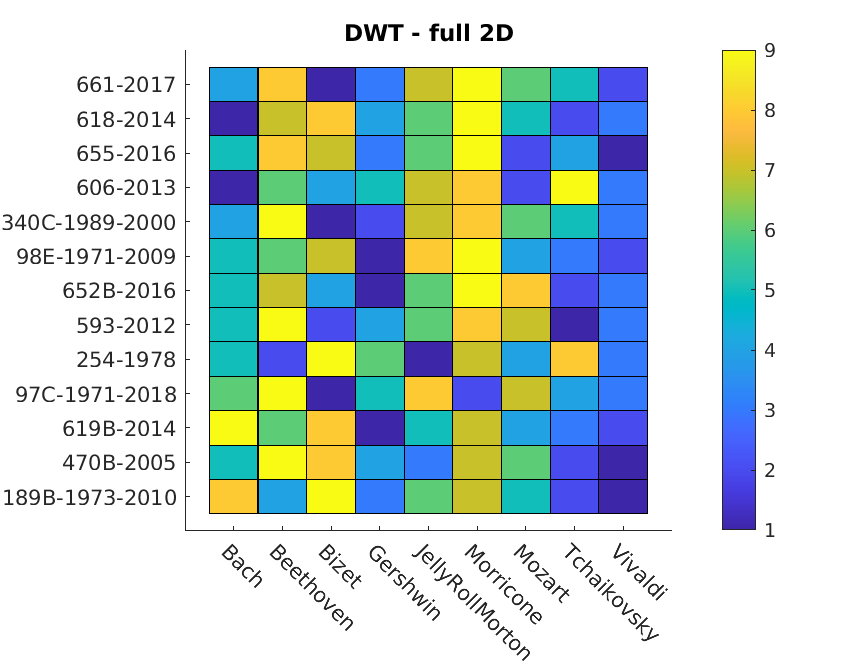}
\end{center}
\caption{\pg Similarity matrices between Morandini's paintings and composers' music, the dark blue {\pg
colour} corresponds
to a greater similarity, the yellow to a lesser similarity. 
The DWT with full 2D transform has been obtained by adopting the strategy
of padding with zero values the coefficients of the image's transform}
\label{fig:dwt-full2d-zeropadding}
\end{figure}

The diagrams of Figure \ref{fig:sintesi1} resume the results obtained.
We notice that, {\pg although} the similarity between the painting and the music tracks depends
on the transform we have used, {\pg the pictures corresponding to
DWT with 1D--unrolling (top--left) and to DFT with both 1D--unrolling and full 2D
transforms (bottom row) are very similar, while DWT with full 2D transform
(top--right) provides a quite different
picture. We guess that this discrepancy 
depends on the inconsistency between the
size of the spectrum of the original image (computed by DWT with full 2D
transform) and that of the audio
tracks (computed by DWT with 1D--unrolling),
and on the fact that we had to align the number of approximation coefficients of the image with that of the music
tracks. In Section \ref{sec:images} we proposed two strategies to
align the sizes of the image and audio transforms: by padding the image transform
with zero values or replicating the coefficients of the image
sequence until the correct size is reached.

The results shown in the top--right picture of Figure
\ref{fig:sintesi1} have been obtained with the latter strategy, while the
results computed by adopting the former strategy (padding with zero values) 
are shown in Figure
\ref{fig:dwt-full2d-zeropadding}. By comparing these two pictures, we notice
that the replicating strategy provides similarity values that are 
closer to those of the other three approaches than those obtained with padding
with zero values. }

Starting from the data in the diagrams of Figure \ref{fig:sintesi1} we have
drawn up the ranking of the composers: 
{\pg the composer of the musics with greatest similarity
to Morandini's paintings is associated with the smallest value,
the one whose musics feature less similarity} with the largest value. 
{\pg As expected, the rankings associated with DWT with 1D--unrolling and to DFT with both 1D--unrolling and full 2D
transforms coincide, while the ranking associated with DWT with full 2D
transform is different.}

\begin{figure}
\begin{center}
\includegraphics[width=0.45\textwidth]{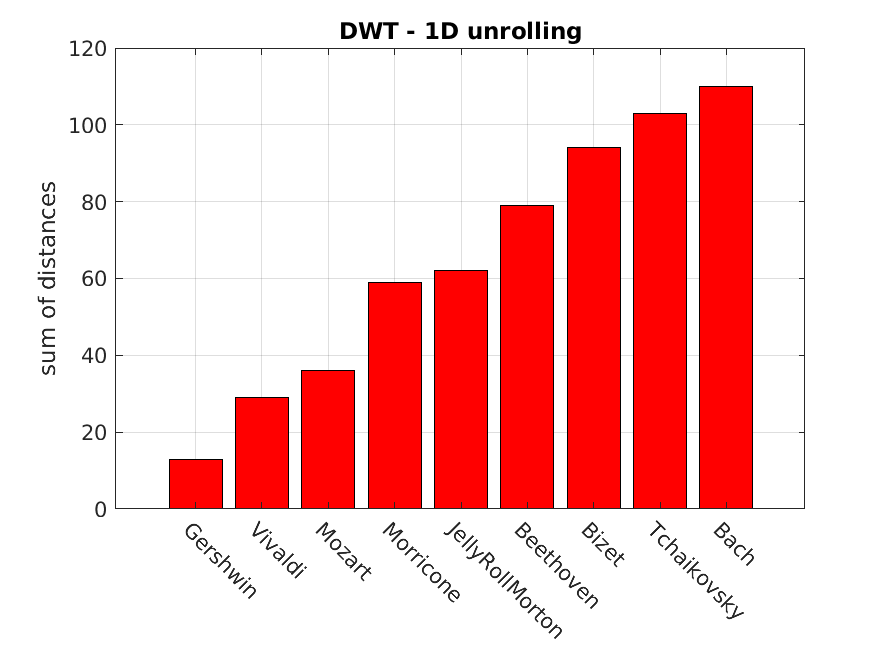}
\includegraphics[width=0.45\textwidth]{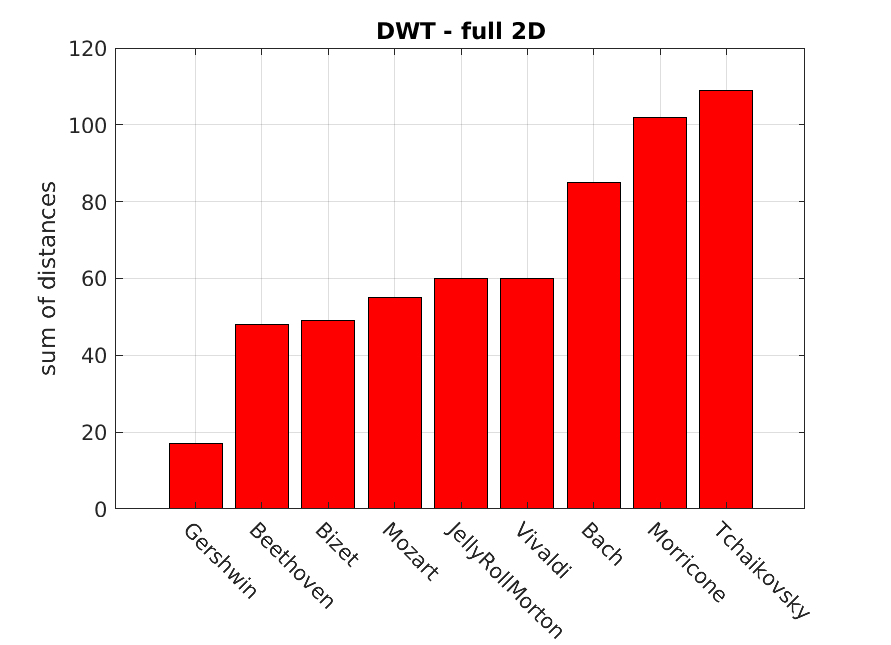}\\
\includegraphics[width=0.45\textwidth]{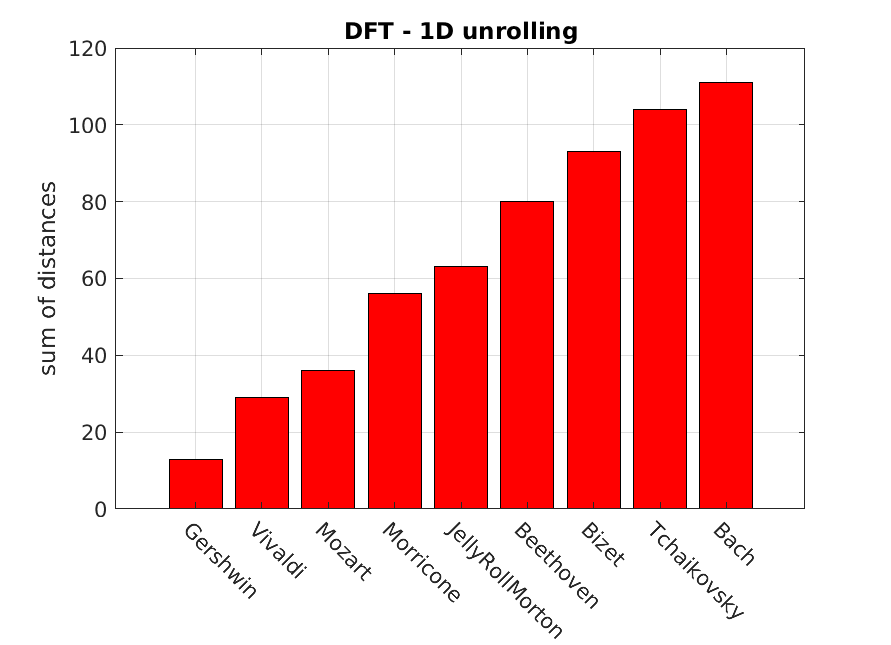}
\includegraphics[width=0.45\textwidth]{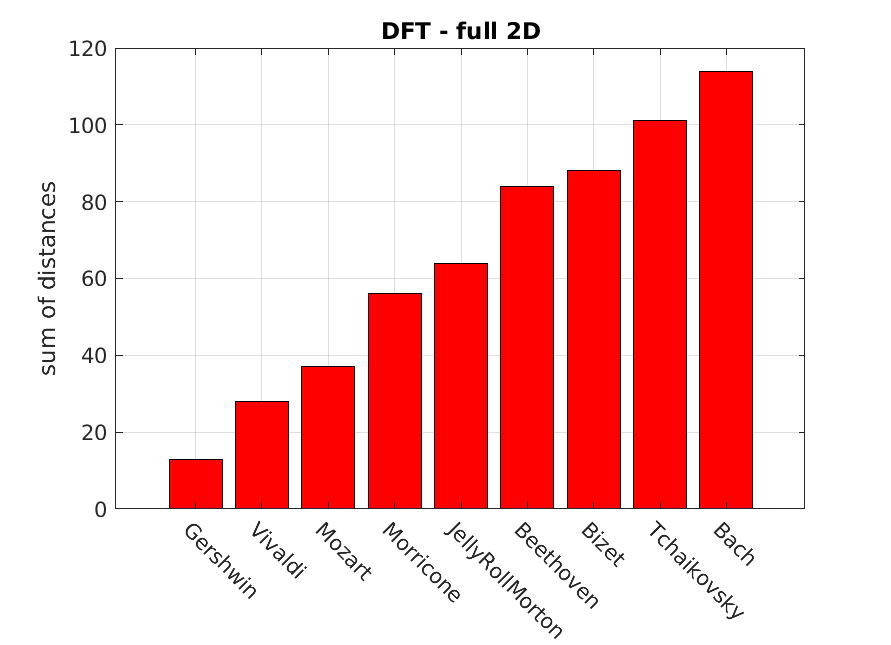}
\end{center}
\caption{Similarity ranking between the {\pg composers' music} and the paintings of
Marcello Morandini: the lower value the better similarity. {\pg The top--right
picture (DWT with full 2D transform) has been obtained by adopting the strategy
of replicating the coefficients of the image's transform}}
\label{fig:sintesi4}
\end{figure}

In Figure \ref{fig:sintesi4} we report the relative histograms: 
{\pg in all cases, the composer (among those previously selected) whose pieces of music
resemble most closely the paintings of Marcello Morandini  is Gershwin (the
relative column in the pictures is that with the most intensive blue colour), 
while
the composers whose pieces show the lower similarity  are:
 Bach and Tchaikovsky when using 
both DWT with 1D--unrolling and DFT (with either 1D--unrolling and full 2D),
Tchaikovsky and Morricone when using DWT with full 2D
transform (the relative columns are those with prevalent yellow colour).}

\begin{sidewaysfigure}
\includegraphics[width=\textwidth]{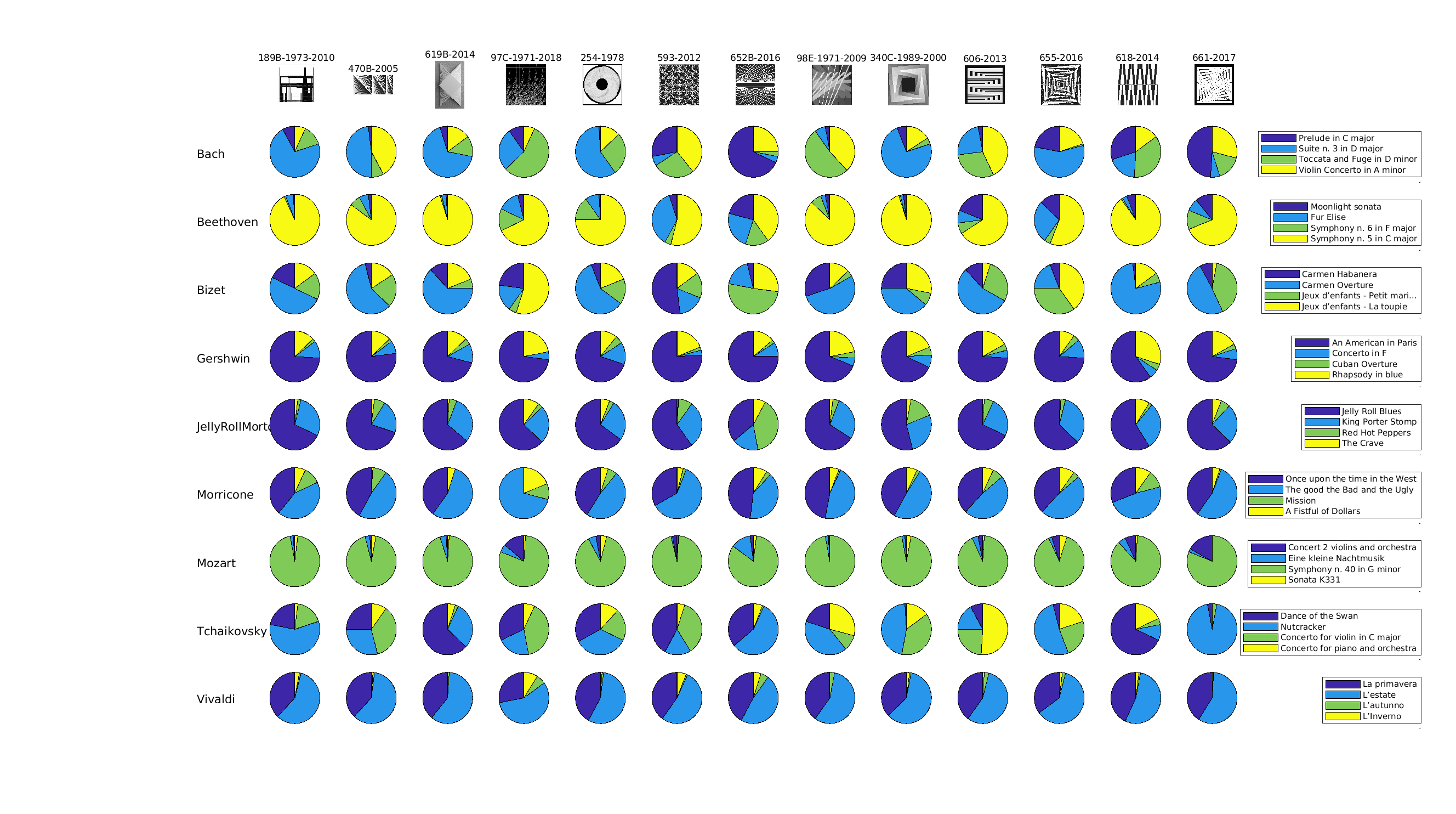}
\caption{\pg The distribution of the percentages $\overline{\alpha}_i$, with
$i=1,\ldots, 4$ when DWT with 1D unrolling strategy is considered. The rows
refer to the composers, the columns to the artworks of maestro Morandini}
\label{fig:pie-dwt1}
\end{sidewaysfigure}

{\pg Finally, in Figure \ref{fig:pie-dwt1} we report the pie--charts provided by
the app showing  the
percentages $\overline{\alpha}_i$ (for
$i=1,\ldots, 4$) when DWT with 1D unrolling strategy is considered. 
We observe that 
among Gershwin's pieces, the one featuring the largest
percentage for all 
Marcello Morandini's paintings is 'An American in Paris'; among Vivaldi's
pieces, it is 'L'Estate, III presto'; among Mozart's pieces, it is
'Symphony 40';  while among Morricone's pieces, it is  'The Good, the Bad, and
the Ugly'. We can interpret such result by saying that 'An American in Paris'
is Gershwin's piece (in our selection) featuring the best similarity with
Marcello Morandini's paintings, 'L'Estate' is Vivaldi's piece (in our
selection) featuring the best similarity with
Marcello Morandini's paintings, and so on.}

{\pg For other composers like, e.g., Bach, Bizet, and Tchaikovsky, we do not
recognize a unique piece that shows the best similarity with all the artworks
of maestro Morandini.}

\begin{figure}
\includegraphics[width=0.9\textwidth]{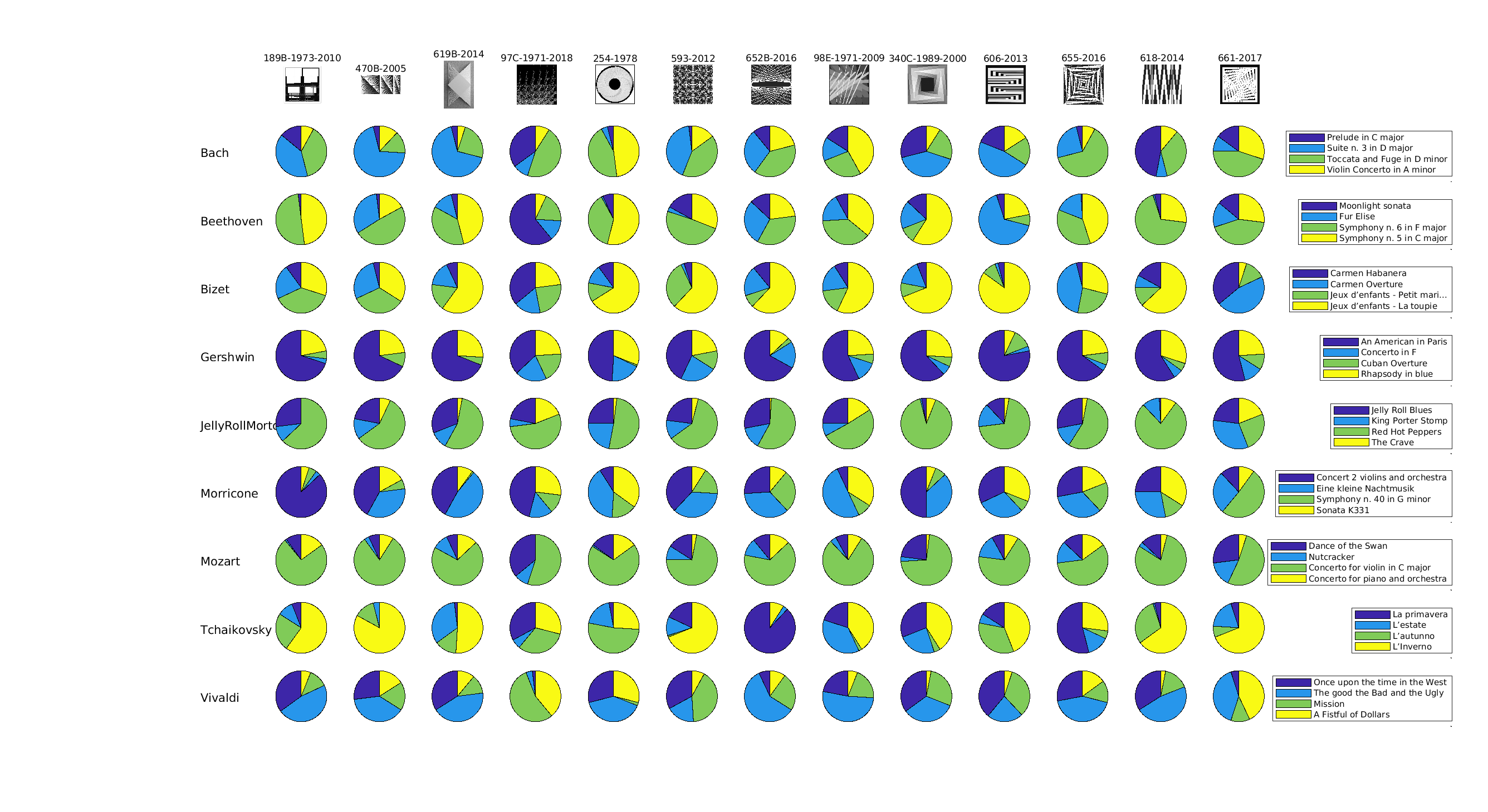}
\includegraphics[width=0.9\textwidth]{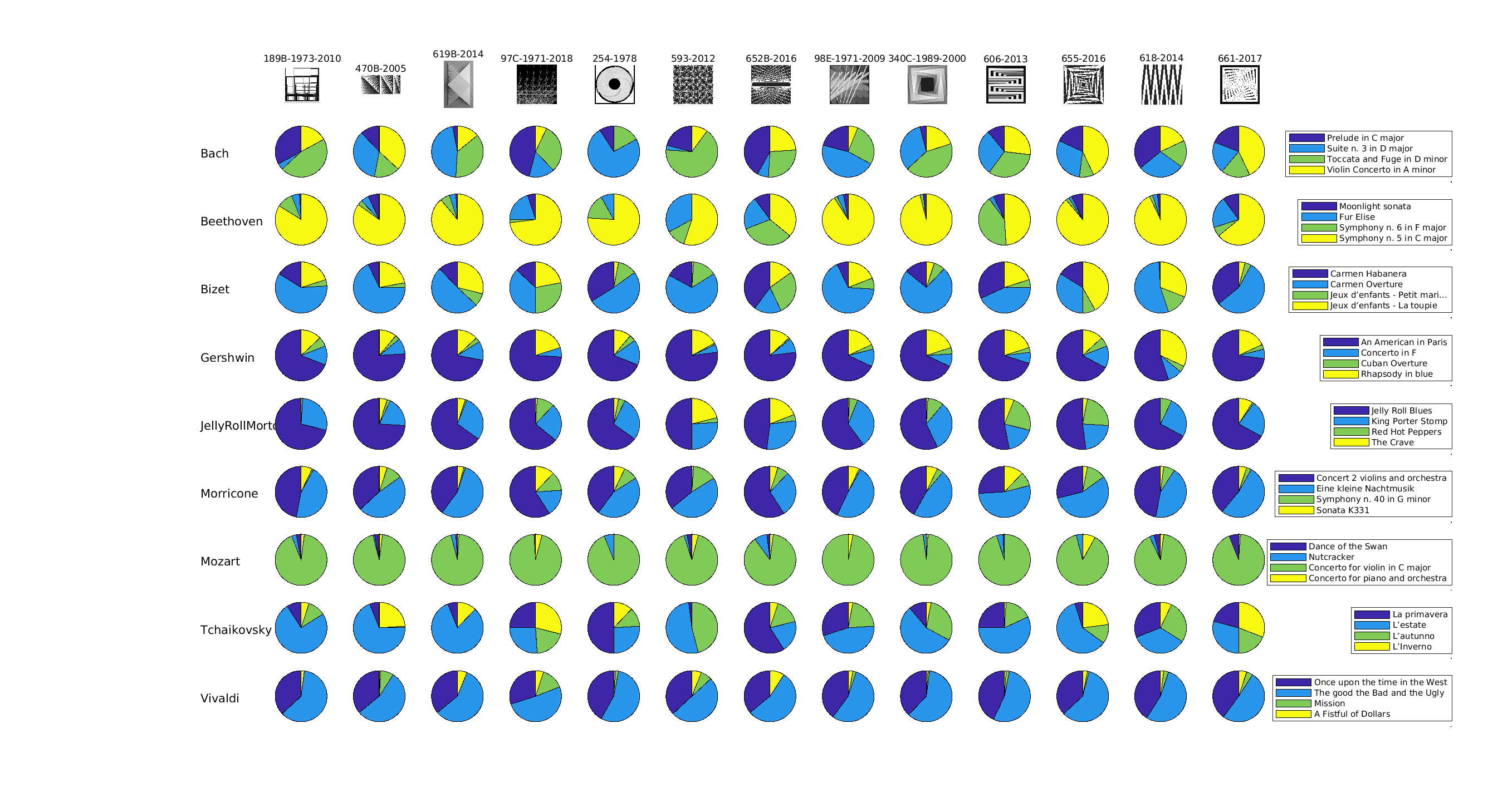}
\includegraphics[width=0.9\textwidth]{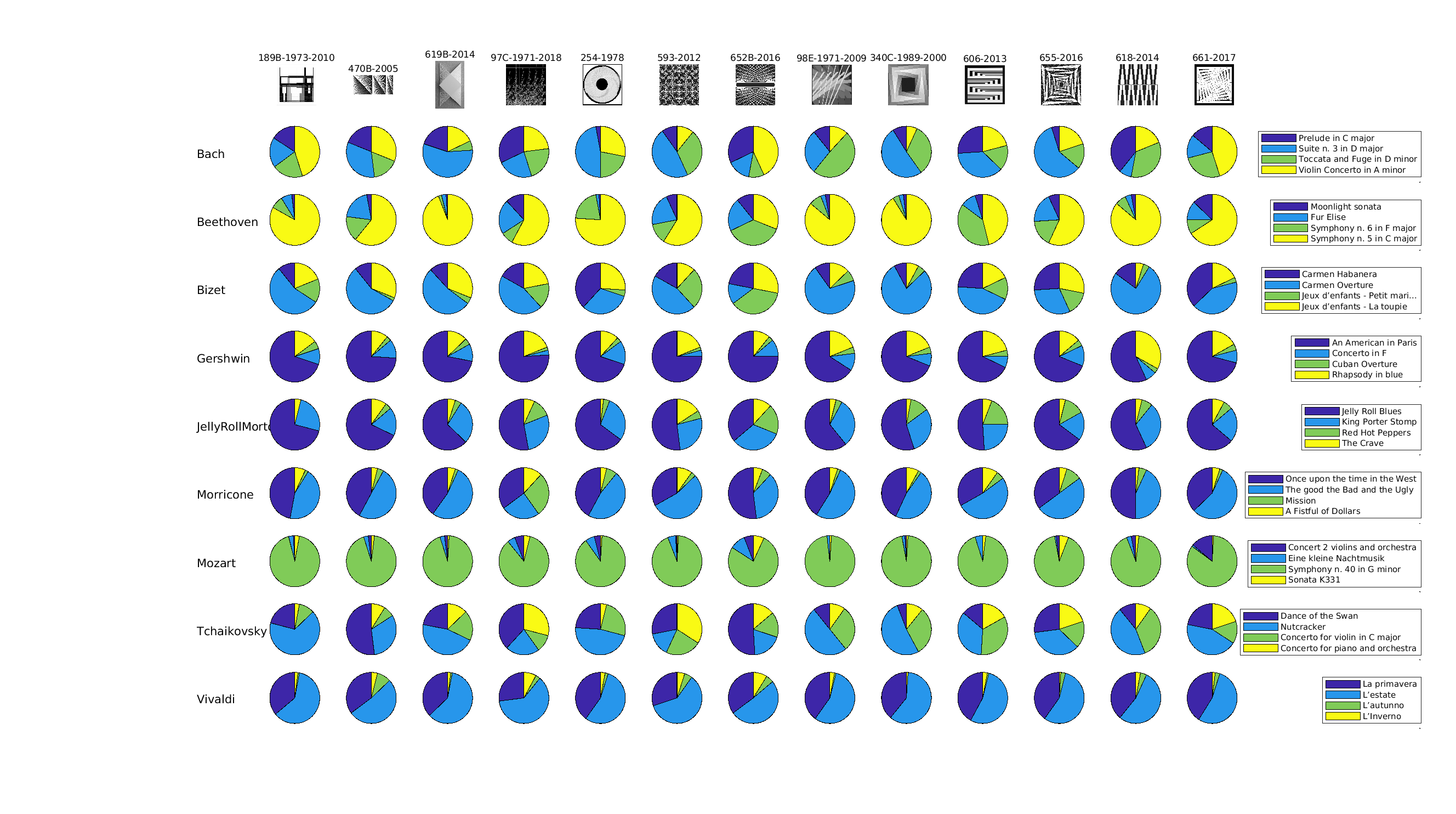}
\caption{\pg The distribution of the percentages $\overline{\alpha}_i$, with
$i=1,\ldots, 4$ when: DWT with full 2D strategy is considered (top), 
DFT with 1D unrolling strategy (center), DFT with full 2D strategy (bottom). The rows
refer to the composers, the columns to the artworks of maestro Morandini. The
order of both composers and paintings, as well as the order of the pieces of
music for each composer, are like in Figure \ref{fig:pie-dwt1}}
\label{fig:pie-other}
\end{figure}

{\pg The percentages computed by our algorithm with the alternative approaches (DWT with
full 2D transform and DFT with both strategies) are shown in Figure
\ref{fig:pie-other}. 
By comparing the pictures in Figure \ref{fig:pie-dwt1} and those in Figure
\ref{fig:pie-other} we {\pg conclude} that, with the exception of DWT with full
2D transform, the transform (either Fourier's and 
wavelets') affects the results, but
does not completely distort them.}

{\pg We conclude this Section by listing
the music tracks and maestro Morandini's artworks we have considered for our
numerical test:}

\noindent
\emph{Bach}: Prelude in C major, BWV 846; 
Suite No.3 in D major, BWV 1068, ``Air''; 
Toccata and Fugue in D minor, BWV 565;
Violin concerto in A minor, BWV 1041;

\emph{Beethoven}: Moonlight Sonata, for B-flat Clarinet and Piano; F\"{u}r Elise; Symphony No. 6 in F major, Op. 68, Pastoral; Symphony No. 5 in C minor, Op. 67, I, Allegro con brio;

\emph{Bizet}: Carmen Habanera; Carmen Overture; Jeux d'Enfants (Petit Mari, Petite Femme); Jeux d'Enfants Op. 22 - II. La toupie;

\emph{Gershwin}: An American In Paris; Concerto In F - 1. Allegro; Cuban Overture; Rhapsody In Blue;
\emph{Jelly Roll Morton}: Jelly Roll Blues;
King Porter Stomp;
Red Hot Pepper;
The Crave;

\emph{Morricone}: Once Upon a Time in the West; The Good, The Bad and the Ugly; Mission; A Fistful of Dollars;

\emph{Mozart}: Concertone for Two Violins and Orchestra K190, Andantino grazioso;
Eine Kleine Nachtmusik K525, I allegro;
Symphony No. 40, K.550;
Sonata K331, 3rd movement;

\emph{Tchaikovsky}: Dance of the Swan;
Nutcracker Dance of the Sugar Plum Fairy;
Violin Concerto in D Major, Op. 35, - I. Allegro moderato;
Piano Concerto No.1 in B Flat minor, Op. 23, Allegro non troppo;

\emph{Vivaldi}: La Primavera, I allegro;
L'Estate, III presto;
L'Autunno, I allegro;
L'Inverno, III allegro;

\emph{Marcello Morandini:}
189B-1973-2010,
470B-2005,
619B-2014,
97C-1971-2018,
254-1978,
593-2012,
652B-2016,
98E-1971-2009,
340C-1989-2000,
606-2013,
655-2016,
412A-2000,
618-2014,
661-2017.

\section{Conclusions and future developments}\label{sec:conclusions}
{\pg To find similarities between paintings and  musics
 we have considered both Fourier and Wavelets
transform, by following a mathematically rigorous approach. Since images and
music tracks are
two--dimensional and one--dimensional data, respectively, we need either to
convert the datum from 2D to 1D before applying the discrete transform, or
to apply the full 2D transform to the image and then adjust it
for a right comparison with the audio tracks' spectra.

Our procedure provides two different results:
from one hand, given a painting and a set of original music tracks, we generate
a new sound track that is the projection of the
painting on the vector space spanned by the original pieces of music. 
On the other hand, we are able to establish which one, among the original sound
tracks used to generate the new music, is the most
similar one to the painting chosen, in terms of intrinsic features.

We found that, with the exception of Discrete Wavelet Transform (DWT) with full
2D transform of the image, the transform (Fourier or wavelet) affects the
results, but does not completely spoil them.

We have implemented our algorithm 
in a Python app available on github and we
have applied our analysis to a selection of 15 artworks of maestro Marcello
Morandini and a total of 36 music tracks chosen among the production of 9
classical, jazz, and modern music composers.

The algorithm we have developed here can be applied to any digital image
representing an artwork and to any set of digital audio files.

At the moment we have considered only 4 music tracks at a time because the
space on our graphical window is limited. However, from the
theoretical point of view we can consider an arbitrary number of music
tracks.

Moreover, the work presented in this paper has been instrumental to the development of 
\emph{RISMapp}, an app for mobile (available on both PlayStore and iStore)  
that, in its preliminary form, lets us \emph{play} some of the artworks of maestro Marcello Morandini
exposed in the Morandini's Foundation in Varese, Italy. 

Future developments will be aimed at tuning and generalizing both the python
and the mobile apps.
}

\section*{Acknowledgments}
Pannello 340C/1989 shown in Fig. \ref{fig:image-pixel} and pannello 593/2012
shown in Fig. \ref{fig:2d-dwt} are two artworks by Marcello Morandini  \cite{Morandini-catalogo2020,Morandini-fondazione2021}.
 The Marcello Morandini Foundation is gratefully acknowledged. A preliminary version of the results of this paper were indeed presented at Fondazione Marcello Morandini in Varese (Italy) during the inaugural Riemann Prize Week celebrations {\tt https://www.rism.it}.
 
The authors report no competing interests to declare.

\bibliographystyle{alpha}
\bibliography{bibl}
 
\end{document}